\documentclass[acmsmall]{acmart}

\AtBeginDocument{%
  \providecommand\BibTeX{{%
    \normalfont B\kern-0.5em{\scshape i\kern-0.25em b}\kern-0.8em\TeX}}}

\renewcommand{\textcolor}[2]{\color{black}#2}

\setcopyright{rightsretained}
\acmJournal{PACMHCI}
\acmYear{2023} \acmVolume{7} \acmNumber{CSCW1} \acmArticle{152} \acmMonth{4} \acmPrice{}
\acmDOI{10.1145/3579628}

\usepackage{graphicx} 
\usepackage{subcaption}

\begin{document}

\title[Toward Supporting Perceptual Complementarity in Human-AI Collaboration]{Toward Supporting Perceptual Complementarity in Human-AI Collaboration via Reflection on Unobservables}

%
%

%
%



\author{Kenneth Holstein}
\authornote{Co-first authors contributed equally to this research.}
\email{kjholste@andrew.cmu.edu}
\affiliation{%
  \institution{Carnegie Mellon University}
   \streetaddress{5000 Forbes Ave}
  \city{Pittsburgh}
  \state{Pennsylvania}
   \country{USA}
}

\author{Maria De-Arteaga}
\authornotemark[1]
\email{dearteaga@mccombs.utexas.edu}
\affiliation{%
  \institution{University of Texas at Austin}
   \streetaddress{110 Inner Campus Drive}
  \city{Austin}
  \state{Texas}
   \country{USA}
}

\author{Lakshmi Tumati}
\email{ltumati@andrew.cmu.edu}
\affiliation{%
  \institution{Carnegie Mellon University}
   \streetaddress{5000 Forbes Ave}
  \city{Pittsburgh}
  \state{Pennsylvania}
   \country{USA}
}

\author{Yanghuidi Cheng}
\email{yanghuic@andrew.cmu.edu}
\affiliation{%
 \institution{Carnegie Mellon University}
 \streetaddress{5000 Forbes Ave}
  \city{Pittsburgh}
  \state{Pennsylvania}
   \country{USA}
}

\renewcommand{\shortauthors}{Holstein, De-Arteaga et al.}

\begin{abstract}
In many real world contexts, successful human-AI collaboration requires humans to productively integrate complementary sources of information into AI-informed decisions. However, in practice human decision-makers often lack understanding of what information an AI model has access to, in relation to themselves. There are few available guidelines regarding how to effectively communicate about \textit{unobservables}: features that may influence the outcome, but which are unavailable to the model.
In this work, we conducted an online experiment to understand whether and how explicitly communicating potentially relevant unobservables influences how people integrate model outputs and unobservables when making predictions. Our findings indicate that presenting prompts about unobservables can change how humans integrate model outputs and unobservables, but do not necessarily lead to improved performance. Furthermore, the impacts of these prompts can vary depending on decision-makers’ prior domain expertise. We conclude by discussing implications for future research and design of AI-based decision support tools.

\end{abstract}





\maketitle


\section{Introduction}
AI-based decision support tools (ADS) are being used to augment human decision-making across a growing range of domains---from predictive risk models used in public services \cite{de2020case,holten2020shifting,kawakami2022improving,levy2021algorithms,saxena2020human}, to AI-based teacher support tools used in K-12 education \cite{an2020ta,chine2022educational,dickler2021using,holstein2020conceptual,ritter2016towards}, to clinical decision support tools used in healthcare \cite{lee2021human,patel2019human,wang2021brilliant,yang2016investigating,yang2019unremarkable}. The sociotechnical design of these systems often relies, whether implicitly or explicitly, on the potential for \textit{human-AI complementarity}. To date, however, scientific and design knowledge remains scarce regarding how we might bring out the best of both human and algorithmic judgment in practice. Recent work offers evidence that AI and human decision-makers can complement each other's capabilities and help to overcome each other's limitations \cite{cheng2022child,de2020case,hemmer2022effect,holstein2022designing}. Yet achieving such synergy in human-AI decision-making is far from guaranteed. For instance, a long line of literature demonstrates that human decision-makers are often either \textit{too skeptical} of useful AI outputs or \textit{too reliant} upon erroneous or harmfully biased AI outputs \cite{buccinca2021trust,dietvorst2015algorithm,green2019principles,lee2004trust,poursabzi2021manipulating}.

In many real-world settings where ADS are employed, humans use AI outputs as just one of several sources of information at their disposal to inform a decision \cite{de2020case,holstein2020conceptual,kawakami2022improving}. 
Thus, humans' ability to successfully integrate information communicated by the ADS with other sources of information is critical to effective decision-making. 
\textcolor{blue}{A central dimension of human-AI complementarity is the often complementary nature of the \textit{information sources} that the human and the AI model each have access to and are able to parse} \cite{de2020case,holstein2020conceptual,kawakami2022improving}. For example, in the context of AI-augmented healthcare, an AI system may have \textcolor{blue}{greater ability to parse through large quantities of time-series data collected via continuous bedside monitoring, while a physician can better perceive changes in patients' emotional state via their physical presentation and by listening to their lived experience \cite{lee2021human,yang2016investigating}. Similarly, in the context of AI-augmented child welfare decision-making, AI models may have access to large quantities of administrative data, while human workers} have access to the rich information communicated during a phone conversation with a caller to a child maltreatment hotline \cite{de2020case,de2021leveraging,kawakami2022improving,kawakami2022care}. 

More generally, consider ADS tools that rely on machine learned prediction models, which estimate an outcome or a probability of an event $Y$ (e.g., a predicted house selling price or the risk of an adverse health outcome), given a set of covariates or features $X$ available to the model. Often, there are additional features $Z$ (often called ``unobservables'' in the machine learning literature \cite{lakkaraju2017selective}) that influence the outcome but are unavailable to the model. In many real-world settings, such as the examples described above, some of these features are available to the human, and it is the hope of those deploying ADS tools that the humans using these tools will be able to complement the AI model's capabilities by integrating this additional information into their decisions. However, prior field research indicates that humans are often unaware of exactly which sources of information are uniquely available to themselves versus an AI model \cite{holstein2019co,kawakami2022improving, kawakami2022care}. Little is known regarding how we might design ADS tools to amplify this form of human-AI complementarity.

In this paper, we investigate the impacts of interventions that prompt people to \textit{reflect on complementary abilities} between themselves and AI models. We focus on \textit{perceptual complementarity}, a form of human-AI complementarity that has been observed in prior work studying real-world human-AI collaborations. In a series of field studies observing how classroom teachers integrate real-time AI recommendations into their decision-making, Holstein et al. \cite{holstein2019co,holstein2022designing} observed that teachers typically cross-checked what the AI system told them about their students with what they were able to see with their own eyes and ears (e.g., student body language and students' own descriptions of the challenges they were facing). In many cases, considering multiple sources of information led teachers to re-interpret the AI recommendations or deem them irrelevant to the actual situation at hand. Similarly, in retrospective data analyses and field observations of AI-assisted decision-making by child maltreatment hotline call screeners, De-Arteaga et al. and Kawakami et al. \cite{de2020case,kawakami2022improving} found evidence that call workers did not indiscriminately adhere to AI recommendations. Instead, call workers were more likely to avoid errors of omission, which correspond to false negative errors, by integrating other sources of information instead of relying on AI outputs alone \cite{de2020case}. However, a series of interview studies with call screeners revealed that they felt uncertain about which information the AI model did or did not have access to, and that they wanted more support from the ADS interface in understanding how exactly this information complemented other information sources at their disposal \cite{kawakami2022improving,kawakami2022care}.

\textcolor{blue}{In real-world decision-making contexts like the education and child welfare examples above, information that is available to humans but unobservable by AI models is often difficult to precisely characterize and externalize without losing important, decision-relevant context and nuance. Indeed, the difficulty of fully capturing such human judgments and perceptions in structured data is one reason why such information may be unavailable to an AI model in the first place. In these settings, human experts often exhibit an impressive ability to effectively integrate rich \textit{implicit} inferences into their decision-making \cite{hemmer2022effect, kawakami2022improving, koedinger2012knowledge, lake2017building, rastogi2022unifying}. In several decision-making contexts, however, unobservables may correspond to structured data. For instance, consider the case of healthcare algorithms, which frequently rely on only a subset of the data available. An algorithm meant to assist physicians in post-cardiac arrest care of comatose patients may predict the likelihood of a positive neurological recovery by relying on EEG signals~\citep{de2019predicting}. Meanwhile, physicians may also have access to complementary information regarding the patient's health (e.g., a series of recorded test results). In either of these scenarios---whether AI unobservables are externalized in structured data or exist primarily in human decision-makers' minds---we expect that promoting reflection on human-AI perceptual complementarity has the potential to improve AI-assisted decision-making.} 
%

\textcolor{blue}{To investigate, we conduct} a 3-condition online experiment to explore how AI-assisted decision-making is impacted by interface prompts that encourage people to reflect on asymmetries in information between themselves and an ADS. In particular, we ask whether and how such prompts might impact (1) how people integrate model outputs with information about features that are unobserved by the model, and (2) whether such prompts improve humans' predictive accuracy. \textcolor{blue}{As a context for this study, we consider an AI-assisted house price prediction task, in which participants are presented with structured data on features of houses \cite{chiang2021you, poursabzi2021manipulating}, some of which are unobserved by the AI model and some of which are more informative than others, and are asked to predict each house's sale price.} 

\textcolor{blue}{Overall, we found that presenting prompts about unobservables can change how people integrate different sources of information. Participants who were shown in-the-moment prompts were more likely to make predictions based on a \textit{combination} of model predictions and their own judgments. Despite this shift, prompting participants about unobservables did not lead to improved predictive performance in the context of our study: human+AI predictions were more accurate than AI predictions across all conditions. Furthermore, we observed that the impacts of these prompts can vary depending on decision-makers' prior domain expertise. For example, among participants who had less prior experience with house price prediction, increasing their overall awareness of the presence of unobservables during a training phase had the effect of \textit{harming} their predictive accuracy. However, this effect was mitigated when these less-experienced participants were also presented with in-the-moment prompts, beyond the training phase, that served as continuous reminders regarding \textit{which specific features} are unobservable to the model.} 

\textcolor{blue}{This work contributes to an emerging body of research in human-computer interaction and CSCW that investigates new ways to help human decision-makers more effectively integrate AI capabilities with their own expertise. Our research builds upon recent work that looks beyond the design of ``AI explanations,'' exploring a broader design space for interactions to support better human-AI decision-making \citep{buccinca2021trust,chiang2021you,poursabzi2021manipulating}. Extending this body of work, we contribute an initial empirical investigation into the impacts of a theoretically promising, yet under-explored interface-level intervention for ADS: in-the-moment prompts that promote reflection on unobservables. We close by discussing key implications of our findings for future work, paying close attention to the complementary roles of different types of unobservables, the volume of information that humans must parse, and the role of the feedback signals available to humans in a given context.}

\section{Background and Related Work}
  The design and evaluation of ADS tools to complement human strengths and improve decision quality has received considerable attention in recent years. In this section we briefly review two bodies of literature that are of particular relevance to our work. 
  
\subsection{Design for human-AI complementarity}

ADS are increasingly used to augment human decision-making across a range of real-world contexts, including education~\citep{an2020ta,dickler2021using,holstein2020conceptual}, healthcare~\citep{lee2021human,wang2021brilliant,yang2016investigating}, social work~\citep{chouldechova2018case,levy2021algorithms,saxena2020human}, and criminal justice~\citep{albright2019if,green2019disparate,kleinberg2018human}. Such systems seek to improve decision quality by leveraging the complementary skills of humans and AI systems. To achieve this, different configurations for dividing labor and integrating human and algorithmic assessments have been proposed. \textcolor{blue}{A stream of work in machine learning has focused on developing automated mechanisms to divide tasks among humans and AI systems (e.g., to ensure that AI systems handle the instances that are most difficult for humans and vice versa), ~\citep{madras2018predict,wilder2020learning,gao2021human,rastogi2022unifying}. 
However, the standard practice in real-world, high-stakes domains maintains decision-making power in the hands of humans: AI outputs are used by a human decision-maker as one of multiple available sources of information.}

Central to the design of these systems is the role of humans' discretionary power. \textcolor{blue}{Typically, the hope is that integrating an AI model as a source of information will improve humans' decision quality, while retaining humans' autonomy and responsibility in a high-stakes decision process. Accordingly, recent field research, conducted in real-world AI-assisted decision-making contexts, provides early evidence that, when human workers are empowered to evaluate and (as appropriate) second-guess AI predictions, this may support more effective and equitable decision-making~\citep{cheng2022child,de2020case,holstein2022designing}. However, empirical results in this area have been mixed, and it remains an open research question how ADS tools can best be designed to foster such human–AI synergy.} In some prior studies, integrations of human and machine intelligence have been shown to be more effective than either humans or AI systems working alone (e.g., \cite{de2020case,holstein2022designing}). Yet in other studies, human–AI collaboration has failed to improve or have even harmed decision quality (e.g., \citep{green2019principles,poursabzi2021manipulating,tan2018investigating}). 
For example, empirical work in the criminal justice domain has shown that humans' discretionary adherence may serve to exacerbate disparities across demographic groups~\citep{albright2019if,stevenson2021algorithmic}, while findings in the child welfare domain have shown that human discretion may mitigate disparities when compared to AI assessments in isolation~\citep{cheng2022child,fogliato2022case}. 

\textcolor{blue}{To date, scientific and design knowledge remains scarce regarding what factors facilitate or hinder complementary human-AI performance. To investigate, a line of research in human-computer interaction and CSCW has begun to explore the design space of interactions that can support \textit{humans} in more effectively integrating AI capabilities with their own strengths as human experts. For example, several recent experimental studies have investigated the impacts of interfaces that present human decision-makers with in-the-moment explanations for specific predictions or recommendations from an AI model. Recent results have shown that, contrary to researchers' intuitions, presenting such explanations can often backfire, encouraging humans to over-rely on AI outputs even in the presence of large errors that they may have otherwise been able to notice (e.g., \citep{bansal2021does,poursabzi2021manipulating}).}

\textcolor{blue}{Moving beyond the design of ``AI explanations'', recent work has begun to explore a broader design space for interactions to support better human-AI decision-making. For example, recent experimental results signal potential for relatively simple cognitive cues (e.g., real-time interface prompts warning that a given instance may be out-of-distribution for an AI model) to help foster more effective use of ADS \citep{buccinca2021trust,chiang2021you,poursabzi2021manipulating}. Our research builds upon this prior work, providing an initial empirical investigation into the impacts of a theoretically promising interface-level intervention for ADS: in-the-moment prompts that encourage human decision-makers to reflect on complementary abilities between themselves and AI models.}

\subsection{Algorithmic prediction under unobservables}

In many AI-assisted decision-making settings, humans have access to sources of information that are not available to the AI model, offering a potential source of complementarity. Formally speaking, when ADS are machine learned prediction models that estimate the probability that an event $Y$ will occur given a set of covariates or features $X$, unobservables refer to features $Z$ that influence the outcome $Y$ but are not available to the AI model~\citep{wansbeek2001measurement}. In some cases, a subset of the features $Z$ may be observable to the human. For instance, in the child welfare context an ADS may use administrative data associated with the child and family members to predict the probability that an investigation following a call to a child abuse hotline will result in out of home placement, but the information communicated in the call is only accessible to the call worker~\citep{chouldechova2018case}. 

When important information is not observed by the AI model, there is a risk of unreliable and misestimated predictions~\citep{wansbeek2001measurement}. This risk may be exacerbated in the context of ADS, where labels suffer from the ``selective labels problem''~\citep{lakkaraju2017selective}. For example, in the historical data used to train AI models in the child welfare domain, the result of a child welfare investigation is only available if a human decided to the screen in the call~\citep{de2021leveraging}. The selective nature of labels available for training requires the use of sampling bias correction methods that rely on the assumption that there are no features that are unobservable to the AI model but observed by humans~\citep{lakkaraju2017selective}.

Empirical work has shown that even in settings where AI models display overall better performance, humans can outperform the algorithm for instances with unique or complex characteristics~\citep{karlinsky2019automating}. Recent research has also demonstrated the potential of information asymmetry as a source of complementary human-AI performance, showing that humans can successfully integrate contextual information when making use of AI recommendations~\citep{cheng2022child,de2020case,hemmer2022effect,holstein2022designing}. However, in many settings human decision-makers do not have a clear understanding of what information is and is not available to the algorithm~\citep{holstein2019co,kawakami2022improving,kawakami2022care}, which may hinder their ability to integrate contextual information. Furthermore, little is known regarding how to best design ADS to support human decision-makers in effectively integrating across AI outputs and unobservables. In this study, we investigate whether prompting people to reflect on asymmetries in information changes the way in which humans integrate the information available to them, and whether this significantly improves human-AI performance by allowing human decision-makers to make better use of unobservables.

\section{Methods}

In this study, we conducted an online behavioral experiment to understand whether and how explicitly communicating potentially relevant unobservables influences the way people make use of algorithmic assistance when making predictions. Our primary research questions are:
\begin{itemize}
\item \textbf{RQ1:} Does presenting reflection prompts about unobservables change how people integrate model outputs and unobservables when making predictions? If so, how?
\item \textbf{RQ2:} Does presenting reflection prompts about unobservables help people integrate model outputs and unobservables more effectively?
\end{itemize}

\subsection{Task and Dataset Selection}
 To explore these questions, we sought to select a task for which our study population (crowdworkers and participants recruited via social media) might be expected to have relevant prior knowledge and experience. As such, we avoided specialized tasks used in prior studies exploring AI-assisted decision-making, such as judicial decision-making tasks \cite{fogliato2021impact,green2019disparate,lurie2020crowdworkers}. Instead, building upon other recent crowdsourcing study designs (e.g., \cite{chiang2021you,hemmer2022effect,poursabzi2021manipulating}), we chose AI-assisted house price prediction as a setting for this study, for several reasons. First, many people may have relevant knowledge or experience in predicting house prices. In addition, house price prediction models like Zillow's \textit{Zestimate} are widely deployed, meaning that participants may \textit{already} be familiar with AI-assisted decision-making in this setting. Finally, people may find this task interesting, even if they have not themselves purchased a home \cite{chiang2021you,fan2018house,poursabzi2021manipulating}. The housing data used in this study came from the Ames, Iowa Housing Dataset~\cite{de2011ames}.

 \subsection{Participant Recruitment and Compensation}
 We recruited a total of 664 participants via the crowdsourcing platform Prolific and through social media channels (Facebook groups and Nextdoor). All participants were based in the US, and were over the age of 18. On Facebook, we targeted relevant special interest groups such as ``First Time Home Buyers'' and ``Real Estate Investing,'' with the aim of recruiting participants who have significant experience with house price prediction. Similarly, on Prolific, we advertised for participants who have prior experience browsing homes online (e.g., using online platforms such as Zillow or Redfin). Our online survey included a background question to track participants' self-reported level of experience browsing for homes: participants were asked to indicate how much time they have previously spent browsing homes for sale on a 5-point scale ranging from ``None at all'' to ``A great deal''. In our analysis, we categorize participants who report having previously spent ``A lot'' or ``A great deal'' of time browsing homes online as having \textit{higher prior experience}, and the remaining participants as having \textit{lower prior experience}.
 All participants received a base payment of \$6. To encourage participants to carefully use the available information to make accurate predictions during the main phase of the study, each participant was informed that they would receive an additional \$6 payment if their prediction accuracy ranked in the top 10\% of participants.

\subsection{Task Design}
Participants performed a sequence of 24 house price prediction tasks, split evenly across a training phase and a testing phase, with the assistance of a pre-trained model. All participants saw the same set of 12 houses in both the training and testing phases, although the order was randomized across participants. The training phase was intended to familiarize participants with the house price prediction task and the model's capabilities, and to help them learn to calibrate their predictions to the specific, unnamed US town used for this study \cite{chiang2021you,poursabzi2021manipulating}. After reviewing a house's information (described in more detail below), along with a model's predicted sale price, participants were asked to predict the house's \textit{actual} sale price. During the training phase, participants received immediate feedback after making a prediction: they were shown the actual sale price of the house, alongside their own prediction, the model's prediction, and the house's information. 
During the testing phase, participants were tasked with predicting prices for 12 houses without receiving any feedback.


For each house, participants were shown ``Facts and Features'', representing the values for the eight features in the dataset that were most predictive of house sale prices. These included the \textit{year built}, the \textit{type of heating}, the numbers of \textit{full baths} and \textit{half baths}, whether the house had a \textit{paved driveway}, the \textit{zoning classification}, and ratings of the house's \textit{material and finish} and \textit{overall condition}. We chose to present only eight features to ensure that the full set of features would not be unmanageable for participants to scan through during the study~\cite{poursabzi2021manipulating}. \textcolor{blue}{In addition, to minimize the chances that participants would draw upon prior knowledge about house prices in a specific time period or region of the US, participants were not told the actual location and time of sale for each house. Instead, participants were simply told that all houses were located in ``the same US town'' and that all houses ``sold around the same time.''} 

\textcolor{blue}{Following prior online experiments studying AI-assisted house price prediction (e.g., \cite{chiang2021you,poursabzi2021manipulating}), participants had access only to this tabular information, and were not able to view images of the homes.\footnote{See \citet{hemmer2022effect} for a recent exception. In their study, participants also have access to images of houses.} In a real-world deployment setting, we envision that the kinds of interface prompts explored in our study would prompt human decision-makers to reflect on particular unobservables, without necessarily requiring that these unobservables are externalized in structured data. For example, an interface might prompt reflection on their own, internal perceptions of variables that a ADS interface designer expects to be relevant for decision-making, but which lie outside of the set of features available to to the AI model (such as a patient's presentation in a healthcare context, or a student's current motivations and life circumstances in an education context). However, to simulate such settings in our online experimental study, we not only draw participants’ attention to particular unobservables, but also present study participants with explicit values for each unobservable.}

The model prediction for each house was generated by a linear regression model trained on the Ames, Iowa Housing Dataset~\cite{de2011ames}. To simulate a real-world scenario in which we might expect to see complementary predictive performance between humans and a trained model (cf.~\cite{bansal2021does}), we artificially induced unobservables in the model prediction by removing three of the eight features listed above from the information that the model had access to: \textit{rating of material and finish}, \textit{number of full baths}, and \textit{type of heating}. As discussed below, these features were chosen as the unobservables for this study given that they were the most predictive of a house's sales price in isolation. Thus, participants in our study had access to three features that were unobservable to the model. The resulting ``partial'' model used in this study had an accuracy of 69.77\%, compared with an accuracy of 80.63\% for the ``full'' model that includes all three unobservables. Note that participants never interacted with the ``full'' model, which was only used to guide the study design. We wished to minimize the risk that participants could succeed at our prediction task by learning an overly simple rule such as ``always predict a higher price than the model.'' Thus, we randomly sampled the houses shown to participants subject to the constraint that half of the houses within each of the training and testing phases were ones for which including the unobservable features in a trained model would result in a \textit{higher} house price prediction compared with omitting them (i.e., houses for which the full model would make higher predictions than the partial model), and the other half were ones for which including the unobservables would result in a \textit{lower} house price prediction compared with omitting them.

After completing all of the prediction tasks in the testing phase, participants were asked to briefly describe how they believed they had been making their predictions during the study. In particular, participants were invited to share reflections regarding how the model's predictions informed their own predictions, and whether they paid particular attention to certain features of each house.

\subsection{Unobservable Features}
\label{subsec:unobs}
The three features selected as unobservables, which hold the greatest predictive power in isolation, exemplify the heterogeneity in the \textit{types of unobservables} that may be present in an AI-assisted decision-making scenario. 

\begin{itemize}
\item \textbf{Unobservables with complementary predictive power:} The \textit{rating of material and finish} and the \textit{number of bathrooms} complement the information that is captured by the features observed by the model. Taking these features into account improves predictive power, over and above the model's capabilities. Including the \textit{rating of material and finish} during model training yields a 6.23\% increase in accuracy, and including the \textit{number of bathrooms} yields an additional 4.64\% increase.
\item \textbf{Unobservables with minimal complementary predictive power:} The \textit{type of heating} holds less complementary predictive power beyond the information already captured by the model. Including the \textit{type of heating} during model training yields a 0.80\% increase in accuracy, suggesting that much of the signal this feature carries is already accounted for by the model via correlations with features that the model can observe.
\item \textbf{Unobservables with low variation within the data observed by humans:} In addition, although the \textit{type of heating} was highly predictive on the larger dataset upon which the model was trained, this categorical feature exhibited little variation within the smaller sample of houses presented to study participants. During the training phase, 10 out of 12 houses shown to participants had the same type of heating. This property limits what humans can learn about an unobservable's predictive power based on the immediate feedback provided during the training phase. However, participants who have higher prior experience on the prediction task may still be able to leverage their prior knowledge about this feature's influence.
\end{itemize}

\subsection{Experimental Design}
\begin{figure}
  \includegraphics[width=\textwidth]{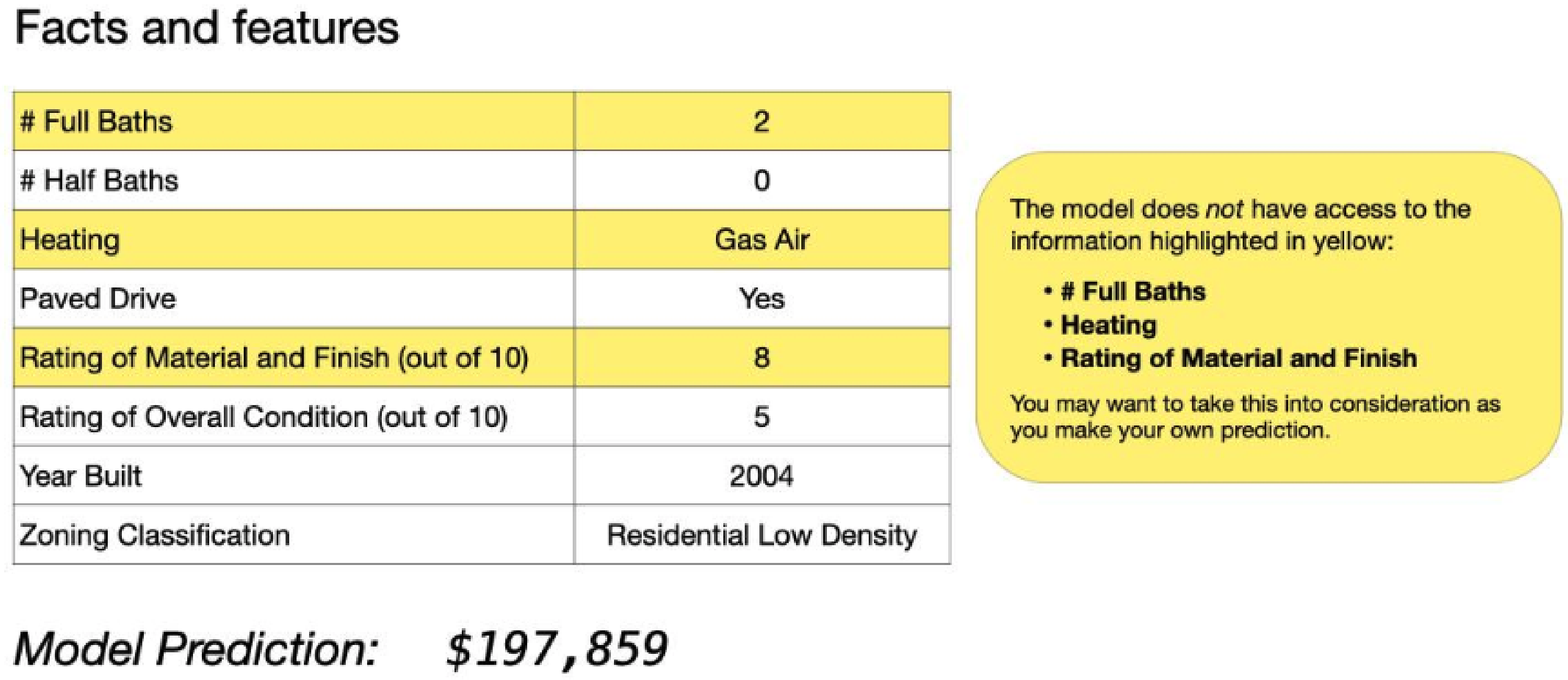}
  \caption{Example of the information presented to participants about a given house. For each house, participants in all conditions were shown a set of eight house features, together with a prediction from a model. In the \textit{Initial + Real-time prompts} condition, shown here, unobservable features were visually highlighted for each house shown, alongside a message indicating that the model does not have access to this information.}
  \label{fig:2}
\end{figure}

Each participant was randomly assigned to one of three experimental conditions. The \textbf{No prompts} condition served as a control condition: participants were not presented with any prompts about unobservables during the study. In the \textbf{Initial prompt} condition, participants received a prompt at the very beginning of the training phase, informing them that the model did not have access to three of the features that they themselves were able to observe, and showing them which features these were (see Figure \ref{fig:2}). However, participants in this condition were not prompted about unobservables again at any point during the study. In the \textbf{Initial + Real-time prompts} condition, in addition to receiving this initial prompt, participants were prompted to reflect on unobservables, as in Figure \ref{fig:2}, every time they were shown house information during the training and testing phases. We included both of these experimental treatments in an effort to tease apart the potential impacts of making participants \textit{aware} of the presence of unobservables, versus \textit{promoting active reflection} on unobservables at the points of prediction and learning from feedback.

\section{Results}

To analyze participants' behavior across conditions, we relied on two central regression models that allowed us to understand how participants integrate model outputs and unobservables across conditions, and whether prediction quality varied across conditions. Let $Y$, $\hat{Y}$, and $\hat{Y}_h$ denote the true selling price of a home, the model's prediction, and  the human's prediction, respectively. Let $Z_{bath}$, $Z_{heat}$, and $Z_{rating}$, denote the three unobservables, corresponding to the number of full baths, type of heating, and rating of material and finish, respectively. Finally, let $C \in \{C_N, C_I, C_{I+R}\}$ denote the experimental condition a participant is assigned to, corresponding to the \textit{No prompts}, \textit{Initial prompt}, and \textit{Initial + Real-time prompts} conditions respectively. 
 
The first model, which we use to answer our first research question below, captures the relationship between participants' prediction and the model's prediction and unobservables, as shown in Equation~\eqref{eq:m1}, where $\psi_j$ denotes random effects for the specific house, $j$.
 \begin{eqnarray}
\hat{Y}_h \sim 1 + \hat{Y} + C + Z_{bath} + Z_{heat} + Z_{rating} + [\hat{Y} + Z_{bath} + Z_{heat} + Z_{rating} ]C + \psi_j
\label{eq:m1}
 \end{eqnarray}
The second model, used to answer our second research question, captures whether and how the magnitude of human's error with respect to the true selling price varies across conditions (see Equation~\eqref{eq:m2}).   
 \begin{eqnarray}
|\hat{Y}_h - Y| \sim 1 + C +  \psi_j
\label{eq:m2}
 \end{eqnarray}

To support interpretation of our statistical results, we also examined the open-text responses that participants provided at the end of the study, capturing participants' own reflections regarding how the model predictions and other information about a house informed their predictions. 
Relevant statistics on participants' open-text responses are used to supplement reporting of our main findings throughout this section, along with relevant quotes from participants. Unless otherwise noted, all statistical results presented correspond to data from the testing phase of the experiment. Regression coefficients presented throughout this section are standardized.

\subsection{RQ1: Effects of prompting on information integration}\label{RQ1}

\begin{figure}
  \includegraphics[width=\textwidth]{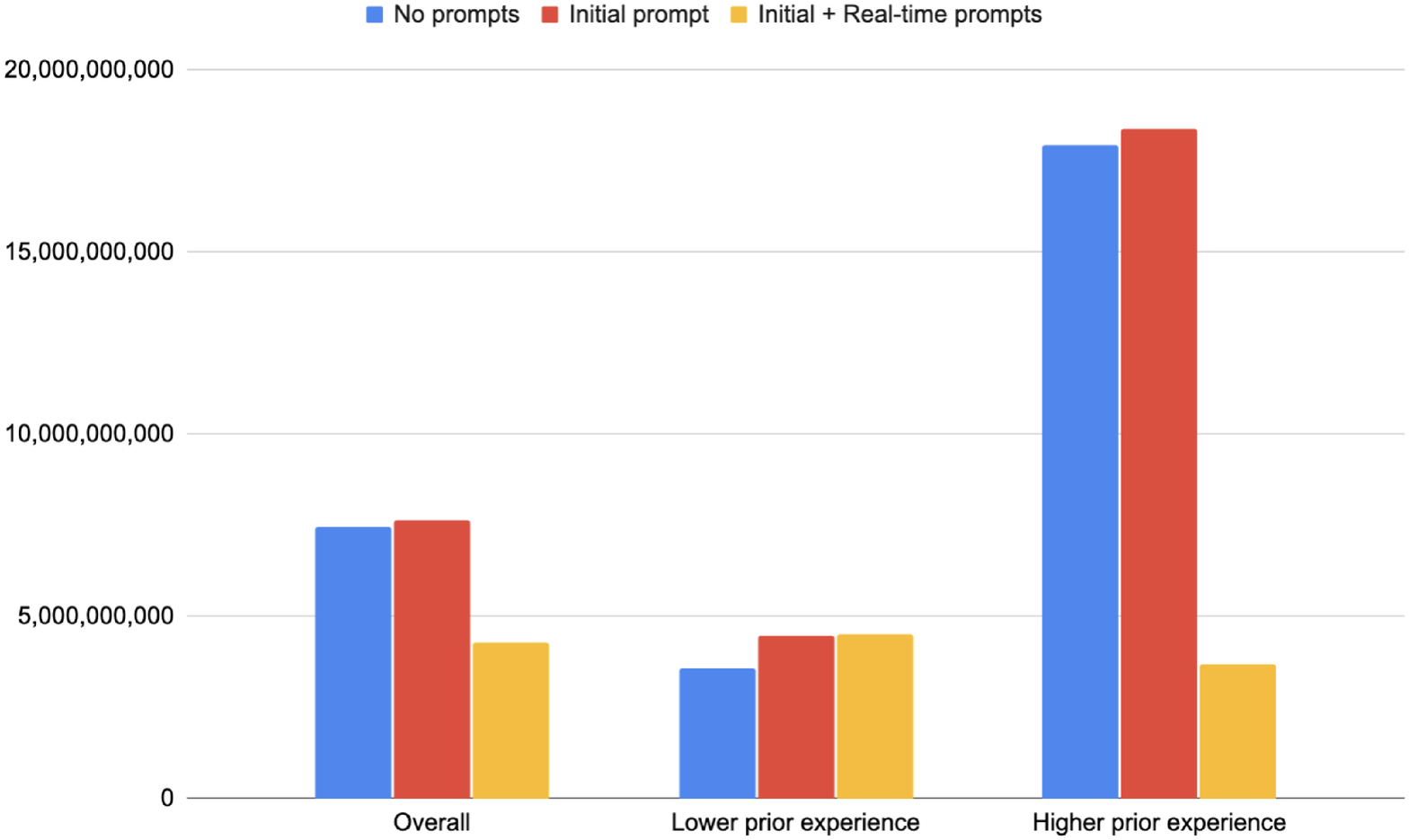}
  \caption{Mean Squared Error (MSE) of regression~\eqref{eq:m1}. In the \textit{Initial + Real-time Prompts} condition, participants' predictions $\hat{Y}_h$ are better explained as a regression over the model's prediction and the unobservables. This is driven by a change in how participants with higher prior experience use the information presented to them.}
  \label{fig:mse}
\end{figure}

We find that presenting prompts about unobservables can change how humans integrate model outputs and unobservables. In particular, when people have constant reminders of unobservables (in the \textit{Initial + Real-time prompts} condition), their predictions are indeed better explained by a combination of the unobservables and the model prediction. This can be observed when analyzing how the Mean Squared Error (MSE) varies across conditions for regression~\eqref{eq:m1}, \textcolor{blue}{shown in Figure~\ref{fig:mse}. This measure indicates how well the participants' predictions $\hat{Y}_h$ can be explained as a regression over the algorithmic prediction and the unobservables. Overall, the MSE is lower in the \textit{Initial + Real-time prompts} condition, reducing to almost half.}

\textcolor{blue}{As shown in Figure~\ref{fig:mse},} this overall reduction in MSE across conditions is entirely driven by a change in participants who reported having higher prior experience browsing for homes (i.e., participants who reported having spent ``A lot'' of time browsing homes online, or greater). When interpreting this result, it is key to consider the difference in magnitudes of MSE between participants with lower versus higher prior experience. Compared with other groups, in the \textit{No prompts} and \textit{Initial prompt} conditions, the MSE is much larger for participants who reported having more prior experience with house price prediction. This indicates that \textbf{in general, participants with higher prior experience may rely less upon the model's prediction and more upon their own integration of features}, including features that are observable to the model, which are not included as independent variables in regression (1). In line with this interpretation, participants with higher prior experience often expressed low confidence in the model in their open-text responses. For instance, a participant with higher prior experience in the \textit{Initial prompt} condition acknowledged that the model was missing information to which they themselves had access, but used this as a justification for ignoring the model's predictions: 
\begin{quote}
    \textit{``The model predictions did not help my own predictions at all, especially since they were missing information. The main things I looked at were number of bathrooms and the material quality.''}
\end{quote}
Similarly, a participant in the \textit{No prompts} condition wrote: 
\begin{quote}
\textit{``I felt the model predictions were completely off and started to ignore them after the training phase - I felt as though they were making me guess incorrectly. I tried to look closest at the number of full and half baths and the year built instead.''} 
\end{quote}


By contrast, in the \textit{Initial + Real-time prompts} condition, the MSE of regression~\ref{eq:m1} for participants with higher prior experience drops to be on par with that of participants with lower prior experience, meaning that in this condition much more of the variation in participants' predictions can be explained by the model predictions and unobservables. This suggests that, in line with the goal of this intervention, \textbf{continuous prompting about unobservables leads participants with higher prior experience to rely more heavily on a \textit{combination} of the model prediction and the unobservables than they otherwise would}. Open text responses from participants with higher prior experience in the \textit{Initial + Real-time prompts} condition support this interpretation. For instance, one participant in this condition described how they integrated their own prior knowledge with the model predictions and information about unobservables:
\begin{quote}
\textit{``I started with the model predictions and assumed the model is taking an average stance (1 or 2 full baths, average material quality) on the unknown variables. I then increased or decreased the price based on those variables and my knowledge of how they affect home prices.''}
\end{quote} 
Another participant similarly described a prediction strategy that used the model prediction as a starting point, and then adjusted based on their knowledge of the unobservables: 
\begin{quote}
\textit{``[I used the] model prediction as a starting base, looked at number of full bathrooms and the overall material rating. If the material rating was 7+ and more than one full bathroom I predicted price to be over the model prediction.''}
\end{quote} 

\begin{table}[!htbp] \centering 
  \caption{Results for regression (1): Relationship between participants' predictions (dependent variable) and model predictions and unobservables (covariates). Standard errors are shown below each coefficient estimate. Significance levels are indicated as: $^{*} ($p$<$0.05), $^{**} ($p$<$0.01), $^{***} ($p$<$0.001).} 
  \label{tab:coef_unobs}
  \resizebox{0.95\textwidth}{!}{
\begin{tabular}{p{25mm}p{25mm}p{37mm}p{40mm}}
\\[-1.8ex]\hline 
\hline \\[-1.8ex] 
 & \multicolumn{3}{c}{\textit{Dependent variable: $\hat{Y}_h$ (human prediction)}} \\ 
\cline{2-4} 
\\
 & \multicolumn{1}{p{25mm}}{Overall}  &
 \multicolumn{1}{p{37mm}}{ Lower prior experience} &  \multicolumn{1}{p{40mm}}{Higher prior experience} \\ 
\hline \\[-1.8ex] 
$C_I$ & \textbf{0.064}$^{***}$ & \textbf{0.093}$^{***}$ & -0.022 \\ 
  & (0.023) & (0.216) & (0.065) \\ 
  & & \\ 
$C_{I+R}$ &   -0.009 & -0.010 & -0.059 \\ 
  & (0.022) & (0.021) & (0.061) \\ 
  & & \\
\hline \\[-1.8ex] 
$\hat{Y}$ & \textbf{0.381}$^{***}$ & \textbf{0.374}$^{***}$ & \textbf{0.400}$^{***}$ \\ 
  & (0.073) & (0.077) & (0.094) \\ 
 & & \\ 
$Z_{bath}$ & \textbf{0.245}$^{***}$ & \textbf{0.251}$^{**}$ & \textbf{0.230}$^{**}$ \\ 
  & (0.053) & (0.056) & (0.068) \\ 
  & & \\ 
$Z_{heat}$ & -0.067 & -0.081 & -0.027  \\ 
  & (0.056) & (0.060) & (0.073) \\ 
  & & \\ 
$Z_{rating}$ & \textbf{0.223}$^{**}$ & \textbf{0.217}$^{*}$ & \textbf{0.241}$^{**}$ \\ 
  & (0.065) & (0.069) & (0.084) \\ 
  & & \\ 
\hline \\[-1.8ex] 
$C_I * \hat{Y}$ & -0.058 & -0.024 & -0.170 \\ 
  & (0.048) & (0.045) & (0.135) \\ 
  & & \\ 
$C_{I+R} * \hat{Y}$ & -0.047 & -0.050 & -0.043 \\ 
  & (0.046) & (-0.044) & (0.125) \\ 
  & & \\ 
$C_I * Z_{bath}$ & 0.038 & 0.053 & -0.022 \\ 
  & (0.035) & (0.032) & (0.098) \\ 
  & & \\ 
$C_{I+R} * Z_{bath}$ &  0.007 & -0.012 & 0.055 \\ 
  & (0.034) & (0.032) & (0.091) \\ 
  & & \\ 
$C_I * Z_{heat}$ & 0.021 & 0.022 & 0.025 \\
  & (0.037) & (0.035) & (0.105) \\ 
  & & \\ 
  $C_{I+R} * Z_{heat}$ & 0.035 & 0.055 & -0.018 \\ 
  & (0.036) & (0.035) & (0.098) \\ 
  & & \\ 
  $C_I * Z_{rating}$ & 0.033 & 0.005 & 0.133 \\ 
  & (0.043) & (0.035) & (0.012) \\ 
  & & \\ 
    $C_{I+R} * Z_{rating}$ & 0.073 & \textbf{0.093}$^{*}$ & 0.019 \\ 
  & (0.042) & (0.035) & (0.112) \\ 
  & & \\ 
\hline \\[-1.8ex] 
Observations & \multicolumn{1}{p{25mm}}{7,968} &  \multicolumn{1}{p{25mm}}{5,904} &  \multicolumn{1}{p{25mm}}{2,064} \\ 
\hline 
\hline \\[-1.8ex] 
\end{tabular}}
\end{table} 

In Table~\ref{tab:coef_unobs}, we also observe a significant interaction between the \textit{Initial + Real-time Prompts} condition and the most informative unobservable for participants with lower prior experience. This suggests that \textbf{continuous prompting about unobservables leads participants with lower prior experience to make greater use of unobservables when making their predictions}. We did not observe comparable effects of presenting a single, initial prompt about unobservables at the beginning of the training phase, in the \textit{Initial Prompt} condition.

Despite these differences across conditions in the way participants integrated the information presented to them, Table~\ref{tab:coef_unobs} shows that \textbf{participants across \textit{all} conditions and levels of experience learned to make use of two of the unobservable features}: the \textit{rating of material and finish} and the \textit{number of bathrooms}. In line with these results, we observed that across all conditions, open-text responses from 80\% of participants explicitly referenced taking into account one or more of the unobservable features. Participants' open-text responses indicated that some participants in our control condition learned to leverage unobservables in order to adjust for limitations of the model's prediction. For example, a participant in the \textit{No prompts} condition wrote: 
\begin{quote}
    \textit{``I noticed the model was occasionally spot-on but also very far off sometimes. So I used it as sort of a middle ground. The number of full bathrooms and year built were the main factors I used to determine the home's value.''}
\end{quote}
 Another participant in the \textit{No prompts} condition similarly stated that: \textit{``If the [ratings] were above 6, I would predict a higher price than the model prediction and vice versa. I also looked at the year the house was built and used that with the other information to determine a price.''} Across conditions, 62\% of participants explicitly stated that they took the \textit{number of bathrooms} into account when making their predictions, 51\% said they accounted for the \textit{rating of material and finish}, and 4\% said they took the \textit{type of heating} into account. 

In interpreting this set of results, it is important to note that participants in our study were presented with a relatively small set of eight features. Had participants been presented with a significantly larger number of total features, making it more difficult and time-consuming to manually inspect all of them, it is possible that participants would have been less likely to pick up on the importance of these unobservables without the type of explicit prompting provided in our experimental conditions. This, combined with the presence of feedback during the training phase, is likely to have played an important role~\cite{lejarraga2021experimental}. 
The interaction of feedback and prompts may have 
helped participants calibrate their use of unobservables. For participants with lower prior experience, we observe a significant interaction between the unobservable with low variation and minimal complementary predictive power (\textit{type of heating}), and the \textit{Initial + Real-time Prompts} and \textit{Initial} conditions during training (see Appendix). This suggests that presenting prompts stating that the model does not observe this information nudges participants to try to make use of it, if they lack prior experience with house price prediction that may guide them to act otherwise. However, this effect disappears in the test phase, presumably after they have learned from feedback that this feature is not useful. Notably, across conditions, participants with higher prior experience quickly honed in on using the model value and the two most informative unobservables, relying primarily on these features to inform their predictions (see Appendix).

    

\subsection{RQ2: Effects of prompting on human predictive accuracy}
\label{subsec:res_rq2}

Overall, we found that \textbf{participants were able to successfully integrate across the information presented in order to make more accurate predictions compared with the model in isolation}. Across all conditions, participants' predictions were \$17,840.84 USD closer to a house's true sales price than the model's predictions on average (p<0.001).
Broken down by condition, this gap in mean absolute error was \$17,465.84 in the \textit{No Prompts} condition, \$15,559.71 in the \textit{Initial prompt} condition, and \$20,225.97 in the \textit{Initial + Real-time prompts} condition.

However, as shown in Table~\ref{tab:disparities}, we did not observe significant improvements in human predictive accuracy over the \textit{No prompts} condition in either the \textit{Initial prompt} condition or the \textit{Initial + Real-time Prompts} condition. \textbf{Although continuous prompting about unobservables did affect how participants integrated the information presented to them, this did not translate into better overall predictive performance.} 
%
%
%
%
It may be that we do not observe significant improvements in predictive performance across conditions because, as discussed in Section~\ref{RQ1}, participants across all conditions and levels of experience learned to make use of the unobservables in our study. As we will discuss in the next section, participants may be less able to do so without the aid of prompts when presented with a larger total number of features. Future work is needed to explore whether the impacts we observe on participants' information integration may translate to improved predictive performance in other contexts.

Interestingly, we observe that the \textit{Initial prompt} condition led to an increase in prediction error among participants who had lower prior experience on the prediction task. We interpret this finding in light of our prior observations that (1) this group of participants made greater use of unobservables when making predictions in the \textit{Initial + Real-time prompts} condition but not in the \textit{Initial prompt} condition (Section~\ref{RQ1}), and (2) the \textit{Initial prompt} condition had a significant overall impact on the magnitude of participants' predictions among those with lower prior experience, but not those with higher prior experience (see Table~\ref{tab:coef_unobs}, Row 1). Taken together, these findings suggest that \textbf{simply increasing participants' awareness of the presence of unobservables during training may lead those with less experience astray}, influencing them to adjust their predictions in ways that harm their predictive accuracy compared with those in other conditions. By contrast, \textbf{continuous reminders regarding which specific features are unobservable to the model appear to mitigate this effect}, in the \textit{Initial + Real-time prompts} condition.

%
%

\begin{table}[!htbp] \centering 
  \caption{Results for regression (2): relationship between participants' absolute prediction error (dependent variable) and experimental condition (covariates). Standard errors are shown below each coefficient estimate. Significance levels are indicated as: $^{*} ($p$<$0.05), $^{**} ($p$<$0.01), $^{***} ($p$<$0.001).} 
  \label{tab:disparities}
  \resizebox{0.9\textwidth}{!}{
\begin{tabular}{p{25mm}p{25mm}p{40mm}p{40mm}}
\\[-1.8ex]\hline 
\hline \\[-1.8ex] 
 & \multicolumn{3}{c}{\textit{Dependent variable: $|\hat{Y}_h - Y|$}} \\ 
\cline{2-4} 
\\
 & \multicolumn{1}{p{25mm}}{Overall}  &
 \multicolumn{1}{p{40mm}}{ Lower prior experience} &  \multicolumn{1}{p{40mm}}{Higher prior experience} \\ 
\hline \\[-1.8ex] 
$C_I$ & 0.017 & \textbf{0.035}$^{*}$ & -0.028 \\ 
  & (0.020) & (0.018) & (0.058) \\ 
  & & \\ 
$C_{I+R}$ & -0.030 & -0.009 & -0.087  \\ 
  & (0.019) & (0.018) & (0.054) \\ 
  & & \\
\hline \\[-1.8ex] 
Observations & \multicolumn{1}{p{25mm}}{7,968} &  \multicolumn{1}{p{40mm}}{5,904} &  \multicolumn{1}{p{40mm}}{2,064} \\ 
\hline 
\hline \\[-1.8ex] 
\end{tabular}}
\end{table} 
  
  
\section{Discussion and Future Work}\label{Discussion}

In this work, we have conducted an online experiment to study the impact of interventions that prompt people to reflect on complementary sources of information available to themselves versus an AI model. We found that presenting prompts about information that is unobservable to the model can change how humans integrate integrate different sources of information when making predictions. 
Even in the absence of improvements in predictive performance, it is noteworthy that such prompts can have an impact on human information integration. 
Given the prevalence of unobservables in real-world AI-assisted \textcolor{blue}{decision-making} settings, more work is needed to investigate what other forms of prompting may be helpful and in which settings.

Moreover, a change in information integration may be of practical relevance even if it does not translate into better overall predictive performance. In the machine learning literature, researchers have studied the phenomenon of \textit{predictive multiplicity}: multiple different models (or human-model teams) may yield the same overall performance~\citep{marx2020predictive}, while having very different fairness properties and underperforming for different \textit{subsets} of cases~\citep{chouldechova2017fairer}. Thus, given our finding that prompts about unobservables can shift how humans integrate information, future work should investigate the impacts of such prompts on relevant metrics beyond predictive accuracy. For instance, future work may consider the fairness implications of prompts that invite reflection on complementary information. From a fairness perspective, it is worth noting that a shift in the grounds of decision-making may also have relevance from a procedural justice perspective, insofar as it may ensure the consideration of important sources of information.

In our experiment, these changes in the ways participants integrate information when receiving prompts about unobservables did not translate into an overall improvement in human predictive performance. When interpreting these results, it is important to consider several characteristics of our study. First, the number of features available to humans in this initial study was relatively small (eight features), which means that it was possible for them to scan through all of the available information in a short period of time. Yet in many real-world settings where AI-based decision support tools are used, the total number of model-observable features and relevant unobservables is significantly larger. For example, call workers tasked with screening child maltreatment calls have hundreds of features available to them~\citep{chouldechova2018case,kawakami2022care}. In settings where a larger number of features are present, it is possible that highlighting the pieces of information that are not available to an AI model would have a greater effect in improving their predictive performance. Furthermore, our experiment included a training phase in which participants were provided with immediate feedback on the accuracy of their own predictions and the model's predictions (cf.~\cite{poursabzi2021manipulating}). Combined with the small number of features present, this may have facilitated participants' learning of what information held complementary power~\cite{lejarraga2021experimental}. In domains where decision-makers do not receive immediate feedback~\cite{holstein2019co,kawakami2022improving}, prompts emphasizing unobservables may have a greater effect in enhancing complementarity. Further research is needed to investigate how the number of features present and the availability of such granular feedback may mediate the impacts of interface prompts about unobservables.

The tension between facilitating better information integration and inducing algorithm aversion~\citep{dietvorst2015algorithm} is one that merits further attention. As discussed in Section~\ref{subsec:res_rq2}, the \textit{Initial Prompt} condition led to an increase in prediction error among participants who had lower prior experience with this task. This may indicate that the prompt served to reduce participants' trust in the model, rather than guiding them on how to use the model more effectively. Thus, designers of interventions that communicate models' limitations, in an effort to help decision-makers calibrate their reliance on these models, must pay close attention to the risk of ``overshooting'' with these interventions and thus inducing algorithm aversion. It is worth noting that in our experiment, the apparent induction of algorithm aversion observed in the \textit{Initial Prompt} condition was mitigated when participants were continuously reminded about \textit{which specific features} were unobservable to the model in the \textit{Initial + Real-time Prompts} condition. It is possible that continuous prompting helped participants to scope their skepticism about model predictions appropriately, so that they were still able to benefit from the model's predictive strengths.

Finally, we note that a nuanced view of different \textit{types} of unobservables is fundamental to the design of sociotechnical systems that aim to facilitate human-AI complementarity. In Section~\ref{subsec:unobs} we differentiate between different types of unobservables, highlighting both (1) their complementary predictive power and (2) the possibilities they present to support humans in learning how to leverage them alongside model predictions, based on feedback. The risk of ``redundancy'' of features that are individually predictive but do not complement the information already captured by the model is particularly interesting to consider. Although a feature may be highly predictive of an outcome of interest in isolation, accounting for this feature may not meaningfully improve predictive power, due to correlations with the observable features that the model already takes into account. This points to a risk that has received little attention in the literature on AI-assisted decision-making to date: human decision-makers may be at risk of ``double counting'' when presented with features that are evidently informative in isolation but whose signal is already accounted for by the model via correlations. Interestingly, we do not observe instances of such double counting in our experiments. However, this may be because the feature that has this property in our study is \textit{also} one that exhibits relatively little variation within the set of instances observed by participants---thereby making it easier for participants to learn from feedback that this feature is of little relevance. Thus, we believe the phenomenon of double counting in AI-assisted decision-making merits further research. We note that the challenge of redundant unobservables can be particularly difficult to tackle from a design perspective: when a feature is an unobservable, this is often because it is not encoded in existing datasets. This means that it is not possible to conduct a data-driven diagnosis to determine whether its predictive power is complementary with that of the observed features. Thus, in the absence of strong theoretical or expert domain knowledge, it is difficult to know whether prompting about a given unobservable could risk misleading participants into ``double counting" if the feature is already \emph{indirectly} accounted for. 

\textcolor{blue}{Our work has direct implications for cases in which it is easy to characterize unobservables and communicate about them. For example, in the AI-assisted child welfare context, call workers have structured ways of identifying certain key pieces of information communicated in calls made to child maltreatment hotlines. Research suggests that in this setting there is significant heterogeneity in call workers' awareness that the information communicated in the call is not visible to the algorithm~\citep{kawakami2022improving}. The findings we present suggest that such heterogeneity likely impacts how they integrate ADS recommendations into their decisions, and prompts that bring awareness and reflection around unobservables could help to reduce this variation. However, as discussed above, in various real-world setting, unobservables may be challenging to precisely characterize and externalize without losing important, decision-relevant context and nuance. This is often the case, for example, in healthcare settings where physicians consider patients' presentation and emotional state, or in education settings where teachers consider how a student's current life circumstances may impact their academic performance. As a simplifying assumption in this study, we presented participants with explicit values for each unobservable in a tabular format, mirroring the design of prior studies that have adopted this AI-assisted housing prediction task (e.g., \cite{chiang2021you,poursabzi2021manipulating}). However, further research is needed to better understand the impacts of different forms of prompting in settings where unobservables are not externalized in structured data, and exist only as perceptions in human decision-makers' minds. 
}

\textcolor{blue}{
Importantly, it is not necessary for humans to be able to explicitly characterize and externalize unobservables in order to incorporate them into their decision-making. A vast body of research in psychology and human-computer interaction shows that human experts exhibit an impressive ability to integrate rich \textit{implicit} inferences about the world into their decision-making \cite{hemmer2022effect, kawakami2022improving, lake2017building, rastogi2022unifying}. In fact, encouraging human decision-makers to externalize such complex inferences (e.g., by attempting to explicitly write down the criteria they are using to inform their decisions) can sometimes backfire by driving them towards the use of easier-to-communicate, yet less reliable criteria \cite{koedinger2012knowledge, lake2017building, nosofsky2005procedural}. 
However, it is worth noting that prompting humans to attend to AI unobservables could be a double-edged sword, and further research is needed to understand risks and benefits across a broader range of decision-making tasks and contexts. In practice, human decision-makers' ability to reason about certain unobservables may vary across cases. For instance, a physician's ability to accurately interpret a patient's emotional state may depend on whether they have a shared sociocultural background. In such settings, there is a risk that bringing awareness to such unobservables could exacerbate existing undesirable biases in human decision-making. Thus, future work should investigate the effects of prompting about unobservables in sensitive context, and special attention should be devoted to heterogeneity of these effects across different decision-makers.
}

In sum, we have found that presenting interface prompts to support human decision-makers in reflecting upon information asymmetries between themselves and an AI model can measurably change how they integrate model outputs with model-unobservable features. However, the impacts of these prompts on human information integration and predictive performance are significantly more complex than we had anticipated at the outset of this project. We now hypothesize that---in addition to being influenced by decision-makers' prior task experience and the frequency of prompting---the impacts of such prompts may also be influenced by (1) the total number of features presented to human decision-makers, (2) the types of unobservables that are included within prompts (e.g., the extent to which these unobservables truly complement the information presented by the model), and (3) the affordances available in a given context for humans to \textit{learn} how to integrate unobservables with model outputs (e.g., whether humans have the opportunity to learn via practice with immediate feedback \cite{lejarraga2021experimental}). Future research exploring each of these hypotheses is needed in order to understand how we can help human decision-makers better leverage complementary perceptual abilities in the context of human-AI collaboration. It is our hope that these findings will inform further research and design toward the design of tools that can bring out the best of both human and AI abilities.



\begin{acks}
This work was supported by an award from the  UL Research Institutes through the Center for Advancing Safety of Machine Intelligence (CASMI) at Northwestern University, by the Carnegie Mellon University Block Center for Technology and Society (Award No. 53680.1.5007718), and by Good Systems, a UT Austin Grand Challenge to develop responsible AI technologies.
\end{acks}

\bibliographystyle{ACM-Reference-Format}
\bibliography{references}

\appendix

\section{Appendix}

\begin{table}[!htbp] \centering 
  \caption{Results for regression (1) on participant data from the training phase. Standard errors are shown below each coefficient estimate. Significance levels are indicated as: $^{*} ($p$<$0.05), $^{**} ($p$<$0.01), $^{***} ($p$<$0.001).} 
  \label{tab:coef_unobs_train}
  \resizebox{0.95\textwidth}{!}{
\begin{tabular}{p{25mm}p{25mm}p{37mm}p{40mm}}
\\[-1.8ex]\hline 
\hline \\[-1.8ex] 
 & \multicolumn{3}{c}{\textit{Dependent variable: $\hat{Y}_h$ (human prediction)}} \\ 
\cline{2-4} 
\\
 & \multicolumn{1}{p{25mm}}{Overall}  &
 \multicolumn{1}{p{37mm}}{ Lower prior experience} &  \multicolumn{1}{p{40mm}}{Higher prior experience} \\ 
\hline \\[-1.8ex] 
$C_I$ & 0.013 & 0.001 & 0.045 \\ 
  & (0.025) & (0.029) & (0.048) \\ 
  & & \\ 
$C_{I+R}$ &   0.001 & 0.000 & 0.003 \\ 
  & (0.024) & (0.029) & (0.044) \\ 
  & & \\
\hline \\[-1.8ex] 
$\hat{Y}$ & \textbf{0.287}$^{***}$ & \textbf{0.264}$^{***}$ & \textbf{0.351}$^{***}$ \\ 
  & (0.033) & (0.039) & (0.040) \\ 
 & & \\ 
$Z_{bath}$ & \textbf{0.284}$^{***}$ & \textbf{0.32}$^{***}$ & \textbf{0.182}$^{***}$ \\ 
  & (0.041) & (0.049) & (0.049) \\ 
  & & \\ 
$Z_{heat}$ & \textbf{0.120}$^{**}$ & \textbf{0.146}$^{**}$ & 0.049  \\ 
  & (0.036) & (0.043) & (0.044) \\ 
  & & \\ 
$Z_{rating}$ & \textbf{0.159}$^{***}$ & \textbf{0.145}$^{**}$ & \textbf{0.195}$^{***}$ \\ 
  & (0.032) & (0.038) & (0.040) \\ 
  & & \\ 
\hline \\[-1.8ex] 
$C_I * \hat{Y}$ & 0.038 & 0.064 & 0.045 \\ 
  & (0.029) & (0.034) & (0.048) \\ 
  & & \\ 
$C_{I+R} * \hat{Y}$ & 0.042 & \textbf{0.078}$^{*}$ & 0.003 \\ 
  & (0.029) & (0.034) & (0.044) \\ 
  & & \\ 
$C_I * Z_{bath}$ & 0.005 & -0.034 & 0.113 \\ 
  & (0.036) & (0.043) & (0.070) \\ 
  & & \\ 
$C_{I+R} * Z_{bath}$ & -0.009 & -0.044 & 0.086 \\
  & (0.035) & (0.042) & (0.065) \\ 
  & & \\ 
$C_I * Z_{heat}$ & \textbf{-0.065}$^{*}$ & \textbf{-0.093}$^{*}$ & 0.011 \\ 
  & (0.032) & (0.038) & (0.062) \\ 
  & & \\ 
  $C_{I+R} * Z_{heat}$ & \textbf{-0.068}$^{*}$ & \textbf{-0.102}$^{**}$ & 0.024 \\ 
  & (0.031) & (0.037) & (0.057) \\  
  & & \\ 
  $C_I * Z_{rating}$ & 0.010 & 0.027 & -0.04 \\
  & (0.029) & (0.034) & (0.055) \\ 
  & & \\ 
    $C_{I+R} * Z_{rating}$ & -0.048 & 0.054 & 0.032 \\
  & (0.028) & (0.033) & (0.051) \\ 
  & & \\ 
\hline \\[-1.8ex] 
Observations & \multicolumn{1}{p{25mm}}{7,968} &  \multicolumn{1}{p{25mm}}{5,904} &  \multicolumn{1}{p{25mm}}{2,064} \\ 
\hline 
\hline \\[-1.8ex] 
\end{tabular}}
\end{table}




\end{document}